\documentstyle[11pt,appb,epsf,epsfig,axodraw]{article}

\begin{document}
\font\elevenit=cmti10 scaled\magstephalf

%\begin{flushright}
%BI-TP-97/43 \\
%\end{flushright}

\begin{center}

\vspace*{40mm}

{\Large \bf Applications of the Large Mass Expansion}

\vskip 10mm

J.~Fleischer
\footnote{~E-mail: fleischer@physik.uni-bielefeld.de}
A.V.~Kotikov
\footnote{Particle Physics Laboratory,
Joint Institute for Nuclear Research, 141980, Dubna (Moscow Region), Russia}
\footnote{~E-mail:kotikov@thsun1.jinr.dubna.su}
\footnote{~Supported by Volkswagen-Stiftung under I/71~293 }
and
O.~L.~Veretin
\footnote{~E-mail:veretin@physik.uni-bielefeld.de}
\footnote{~Supported by BMBF under 05~7BI92P(9)}

\vskip 10mm

{\it ~Fakult\"at f\"ur Physik, Universit\"at Bielefeld,
D-33615 Bielefeld, Germany.}

\begin{abstract}

The method of the large mass expansion (LME) is investigated for
selfenergy and vertex functions in two-loop order. It has the technical advantage
that in many cases the expansion coefficients can be expressed analytically.
As long as only one non-zero external momentum squared, $q^2$, is involved also
the Taylor expansion (TE) w.r.t. small $q^2$ yields high precision results
in a domain sufficient for most applications. In the case of only one
non-zero mass $M$ and only one external momentum squared, the expansion
w.r.t. $q^2/M^2$ is identical for the TE and the LME. In this case the 
combined techniques yield analytic expressions for many diagrams, which 
are quite easy to handle numerically. 

\end{abstract}
\end{center}

\vfill

\thispagestyle{empty}
\setcounter{page}0
\newpage

\section{Applications of the presented methods in particle physics }

   Diagrams with one non-zero mass and only one external 
$q^2$ have wide applications in QED and QCD for both selfenergies and vertices.
For applications in the electroweak part of the full Standard Model (SM),
due to the rather large spectrum of masses, in general one has to apply
numerical methods. Often, however, approximations like small fermion masses 
$m_f^2/m_Z^2\sim0$, close vector boson masses $(m_Z^2-m_W^2)/m_Z^2\sim0$ and
large top mass $m_Z^2/m_t^2\sim0$ etc. are considered. In these approximations
the problem is often reduced to diagrams with one non-zero mass, which is
a further reason for the interest in this type of diagrams.
For applications in the electroweak
part of the full SM, two calculations in the two-loop order are of particular 
interest:\\

   a) Z decay into $b\bar{b}$ (bottom, anti-bottom)\\

   The kinematics in this case is $q^2=M_Z^2$ and $p_1^2=p_2^2=0$ (in the $m_b=0$
approximation). 

\vspace{1.5cm}
\begin {figure} [htbp]
\begin{picture}(400,100)

\GCirc(200,100){30}{0.8}

\Photon(200,30)(200,70){3}{5}
\ArrowLine(180,120)(150,150)
\ArrowLine(250,150)(220,120)

\Text(155,135)[]{$p_1$}
\Text(245,135)[]{$p_2$}

\Text(145,145)[]{$b$}
\Text(255,145)[]{$\bar{b}$}

\Text(190,50)[]{$Z$}
\Text(237,50)[]{$q=p_1-p_2$}

\end{picture}
\vspace{-1.0cm}
\caption {Kinematics for the decay $Z \to b\bar{b}$.}

\end{figure}
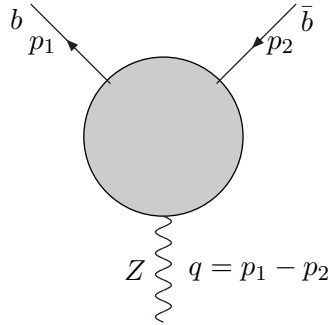

   We mention that the decay $Z \to b\bar{b}$ was considered by a number
of authors \cite{ZT} in the limit of large top mass, i.e. in the leading order
only the top mass was kept non-zero and all other masses put to zero
and even $q^2=m_Z^2=0$. 
This is an example of the above mentioned approximations and here also
the number of diagrams is quite low (of the order of 10).
Recently higher order terms in the large top mass expansion have been
considered \cite{HSS} (see also these proceedings).
In \cite{FLM} an investigation of the precision of the higher orders
in the large mass expansion has been performed for scalar diagrams
occurring in this process
and a repetition of higher orders in the large top mass for the full process
is under consideration.\\

   b) Anomalous magnetic moment (AMM) of the muon ($g-2$)\\

   Here the kinematics is quite different, namely $q^2=0$ (actually $q=0$) and
$p_1^2=p_2^2=m_{\mu}^2$. In fact in this case the problem reduces to the calculation
of two-point functions. In both cases a) and b) we have one external variable only:
$q^2=M_Z^2$ and $p_1^2=p_2^2=m_{\mu}^2$, respectively.\\

\vspace{1.5cm}
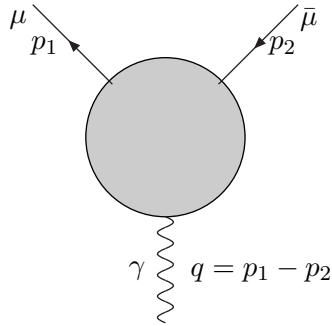
\begin {figure} [htbp]
\begin{picture}(400,100)

\GCirc(200,100){30}{0.8}

\Photon(200,30)(200,70){3}{5}
\ArrowLine(180,120)(150,150)
\ArrowLine(250,150)(220,120)

\Text(155,135)[]{$p_1$}
\Text(245,135)[]{$p_2$}

\Text(145,145)[]{$\mu$}
\Text(255,145)[]{$\bar{\mu}$}

\Text(190,50)[]{$\gamma$}
\Text(237,50)[]{$q=p_1-p_2$}

\end{picture}
\vspace{-1.0cm}
\caption {Kinematics for anomalous moment ${(g-2)}_{\mu}$.}

\end{figure}

   In the case of the AMM our first step was dedicated to an
automatic set up of
the contributing Feynman diagrams. For this purpose the package TLAMM 
was developed \cite{TLAMM}. Also the large mass expansion
(LME) \cite{LME} applied to a ``toy model''
for the ${(g-2)}_{\mu}$ was described in \cite{TLAMM}, i.e. for diagrams
of the selfenergy type. In this model 40 diagrams were contributing
to the two-loop AMM of the muon.\\

   In general, however, in the above cases many more, i.e. of the order 
of 1000 diagrams
contribute and naturally it is necessary to investigate diagrams with
various non-zero masses. 

   As was already started with TLAMM, in the full SM it is even more
required to produce the Feynman diagrams automatically. Instead of
40 diagrams, e.g., in the above mentioned ``toy model'' in the full
${(g-2)}_{\mu}$ calculation 1832 diagrams contribute. Such a more general
program, called {\bf DIANA} (for DIagram
ANAlyser) \cite{DIA}, written in C and making use of Nogueira's QGRAF \cite{QGRAF},
was also developed in our group.
It produces FORM \cite{FORM} input for the Feynman diagrams according to the
Feynman rules for further evaluation.

\newpage

\section{Evaluation of scalar diagrams}

   In the following we shortly review the main techniques we applied, namely
the Taylor expansion (TE) \cite{ft} and the LME \cite{LME}. The efficiency of 
both approaches will be compared.

\subsection{The Taylor expansion method}

      Taylor series expansions in terms of one external momentum
squared, $q^2$ say,
were considered in \cite{Recur}, Pad\'{e} approximants
were introduced in \cite{bft} and in Ref. \cite{ft} it was demonstrated
that this approach can be used to calculate Feynman diagrams on their
cut by analytic continuation.
In the case of a three-point function like $Z \to b\bar{b}$ 
we have two independent external momenta in $d=4-2 \varepsilon$ dimensions.
The general expansion of (any loop) scalar 3-point function with its
momentum space representation $C(p_1, p_2)$ can be written as
\begin{equation}
\label{eq:exptri}
C(p_1, p_2) = \sum^\infty_{l,m,n=0} a_{lmn} (p^2_1)^l (p^2_2)^m
(p_1 p_2)^n
\label{2.2}
\end{equation}
where the coefficients  $a_{lmn}$ are to be determined from the given diagram.
Considering $Z \to b\bar{b}$ with
$m_b=0$, i.e. $p^2_1 = p^2_2 = 0$, which is a good 
approximation, only the coefficients $a_{00n}$ are needed.
For the calculation of the Taylor coefficients in general various procedures have
been proposed \cite{ft},\cite{davt},\cite{Tara}. 
These methods are well suited for programming in terms of a formulae manipulating 
language like FORM. Such programs, however, yield acceptable analytic results 
only in cases when not too many parameters (like masses) enter the problem. 
Otherwise numerical methods are needed \cite{JF}, \cite{Mainz}.

In the case of only one non-zero mass and only one external momentum squared,
indeed the case with the least nontrivial parameters, for many diagrams 
analytic expressions for the Taylor coefficients can be obtained. In Sect. 5 we 
present some recent results.

For the purpose of calculating Feynman diagrams in the kinematical domain
of interest it is necessary to calculate them from the Taylor series on 
their cut. This is performed by analytic continuation in terms of a mapping.

Assume, the following Taylor expansion of a scalar diagram or a
particular amplitude is given
$C(p_1, p_2,\dots)=\sum^\infty_{m=0} a_m y^m \equiv f(y)$
and the function on the r.h.s. has a cut for $y \ge y_0$.

 The method of evaluation
of the original series consists in a first step in a conformal mapping
of the cut plane into the unit circle and secondly the reexpansion
of the function under consideration
into a power series w.r.t. the new conformal variable.
A variable often used is
\begin{equation}
\omega=\frac{1-\sqrt{1-y/y_0}}{1+\sqrt{1-y/y_0}}.
\label{omga}
\end{equation}

\begin{figure}[h]
\centerline{\vbox{\epsfysize=45mm \epsfbox{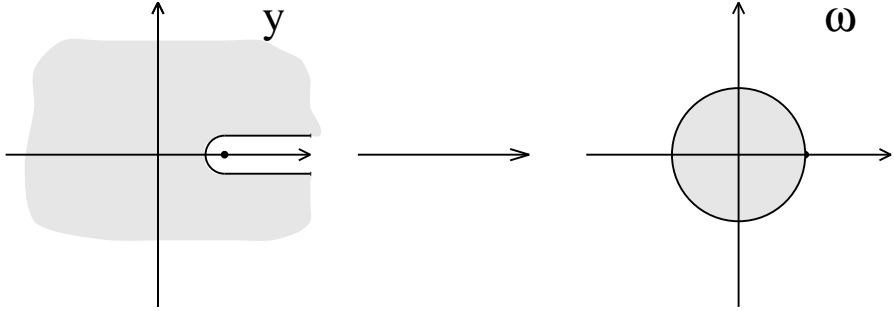}}}
\caption{\label{conf}Conformal mapping of the complex
$y$-plane into the $\omega$-plane.}
\end{figure}

By this conformal transformation,
the $y$-plane, cut from $y_0$ to $+ \infty$, is mapped into the unit
circle (see Fig.\ref{conf}) and the cut itself is mapped on
its boundary, the upper
semicircle corresponding to the upper side of the cut.
The origin goes into the point $\omega=0$.\\

  After conformal transformation it is suggestive to improve the
convergence of the new series w.r.t. $\omega$ by applying the
Pad\'e method \cite{Sha},\cite{BGW}.
A convenient technique for the evaluation of Pad\'e approximations
is the $\varepsilon$-algorithm of~\cite{Sha} which allows one
to evaluate the Pad\'e approximants recursively.

\subsection{Large mass expansion (LME)}

  In particular for the evaluation of diagrams with several different
masses, one of which being large (like the top mass, e.g., $m_t$),
we use the general method of asymptotic
expansion in large masses \cite{LME}. For a given scalar
graph $G$ the expansion in large mass is given by the formula
\begin{equation}
F_G(q, M ,m, \varepsilon) \stackrel{M \to \infty}{\sim }
\sum_{\gamma} F_{G/\gamma}(q,m,\varepsilon) \circ
T_{q^{\gamma}, m^{\gamma}}
F_{\gamma}(q^{\gamma}, M ,m^{\gamma}, \varepsilon),
\label{Lama}
\end{equation}
\noindent
where $\gamma$'s are subgraphs involved in
the LME, $G/\gamma$ denotes shrinking of $\gamma$ to a
point; $F_{\gamma}$ is the Feynman integral corresponding to
$\gamma$; $ T_{q_{\gamma}, m_{\gamma}} $ is the Taylor operator
expanding the integrand in small masses $\{ m_{\gamma} \}$ and
external momenta $\{ q_{\gamma} \}$ of the subgraph $\gamma$
; $ \circ$ stands for the convolution of the subgraph expansion
with the integrand $F_{G/{\gamma}}$. The sum goes over all
subgraphs $\gamma$ which (a) contain all lines with large masses, and
(b) are one-particle irreducible w.r.t. light lines.

   For the $Z \to b\bar{b}$ decay we have $q^2 = M_Z^2$ for the 
on-shell Z's. Fig.4 shows diagrams with two different masses
on virtual lines, one of which is a top. 
$W$ and $Z$ are the gauge bosons with masses $M_W$ and $M_Z$, respectively;
$\phi$ is the charged would-be Goldstone boson (we use the Feynman gauge);
$t$ and $b$ are the t- and b-quarks.

%
%  PICTURES
%
\begin{figure}[h]
\centerline{\vbox{\epsfysize=40mm \epsfbox{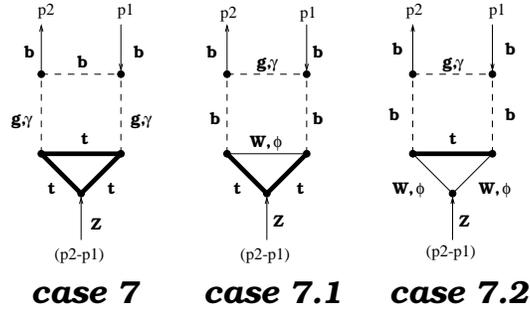}}}
\vspace*{3mm}
\noindent
%Figure 3: Two-loop diagrams with two different masses
%in internal lines arising in the process $Z  \to b \overline{b}$.
\caption{\label{diagrams} Two-loop diagrams with two different masses
in internal lines arising in the process $Z  \to b \overline{b}$.
The notation for the diagrams is chosen according to \cite{ft}.}
\end{figure}

\vspace*{0cm}
\begin{figure}[h]
  \centerline{\vbox{\epsfysize=55mm \epsfbox{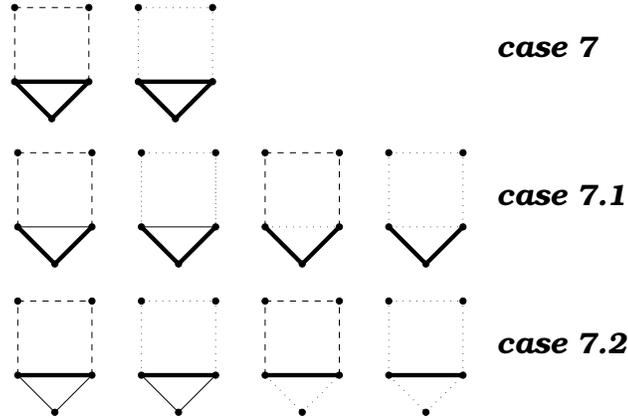}}}
  \vspace*{3mm}
  \noindent 
                                %Figure 2: The structure of the LME, see explanations in the text.\\
                                %\end{figure}
  \caption{\label{expansion} The structure of the LME,
    see explanations in the text.}
\end{figure}

                                %\newpage
~\\
Applying the method of the LME, $M_W$ and $M_Z$ are considered
as small. In the framework of this expansion
contributions from additional subgraphs are to be taken into account
together with the Taylor expansion of the initial diagrams w.r.t. external
momenta and light masses.
These are shown in Fig.5. Bold, thin and
dashed lines correspond to heavy-mass, light-mass, and massless
propagators, respectively. Dotted lines indicate the lines omitted
in the original graph $\Gamma$ to yield the subgraph $\gamma$.
$\Gamma/\gamma$ (see (\ref{Lama})) then consists out of all the dotted lines
after schrinking $\gamma$ to a point.

These subgraphs restore the analytic 
properties of the initial diagrams (like logarithmic behaviour).
In other words, the Taylor expansion of the initial diagrams produces
extra infrared singularities which are compensated by singularities of the
additional subgraphs so that only the singularities of the original diagrams 
survive.

At this point it also becomes clear what the difference is between the
small-$q^2$ expansion and what is called here the LME:
in the former case we assume all masses large, i.e. 
$q^2 \ll M_W^2,M_Z^2,m_t^2$ while in the latter case only $m_t$ is considered
as large and all other parameters small, i.e. $q^2,M_W^2,M_Z^2 \ll m_t^2$.
In this sense both methods are LM expansions. The technical
advantage of the second method is, however, that only bubbles with one mass occur, 
which can be expressed in terms of $\Gamma$ functions, while in
the case of the small $q^2$ expansion bubbles with different masses are involved,
which are much more difficult to evaluate, in fact only numerically. Of course 
also the number and structure of the subgraphs is different in the two cases.

                                %
                                %  GRAPHICS'
                                %

                                %\begin{center}
\begin{figure}[h]
  \vspace*{-1.5cm}
  \hspace*{2cm}\vbox{
    \raisebox{6.0cm}{\makebox[0pt]{\hspace*{-2cm}$m_t^4\mbox{\rm Re}J_{7.1}$}}
    \raisebox{1.5cm}{\makebox[0pt]{\hspace*{18.5cm} $q^2/m_Z^2$}}
    \epsfysize=84mm \epsfbox{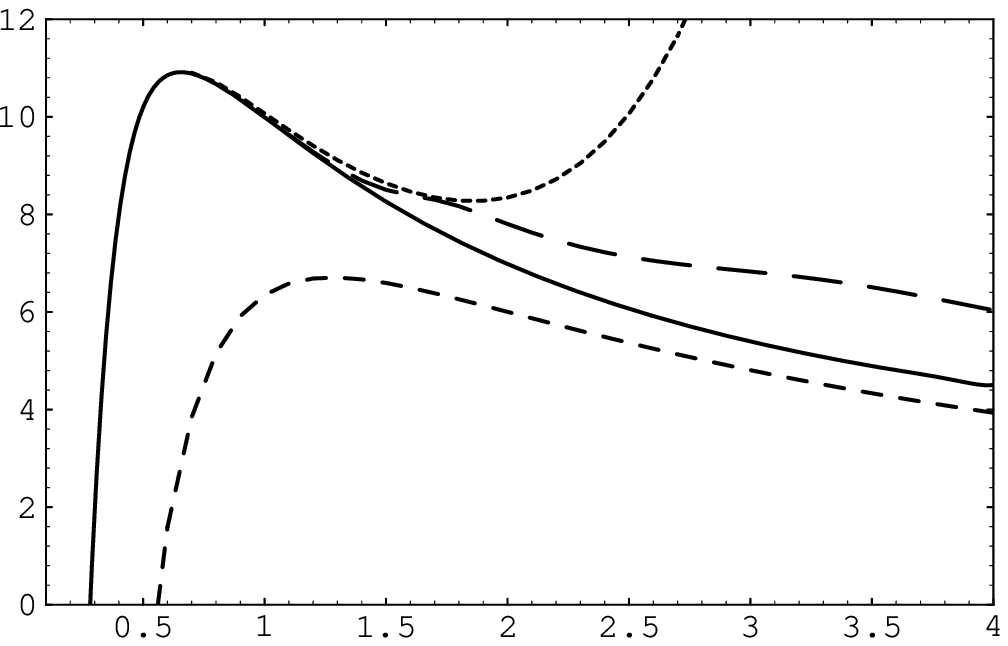}
    }\\
  \vspace*{-2.8cm}
  \hspace*{2cm}\vbox{
    \raisebox{6.0cm}{\makebox[0pt]{\hspace*{-2cm}$m_t^4\mbox{\rm Re}J_{7.2}$}}
    \raisebox{1.5cm}{\makebox[0pt]{\hspace*{18.5cm} $q^2/m_Z^2$}}
    \epsfysize=84mm \epsfbox{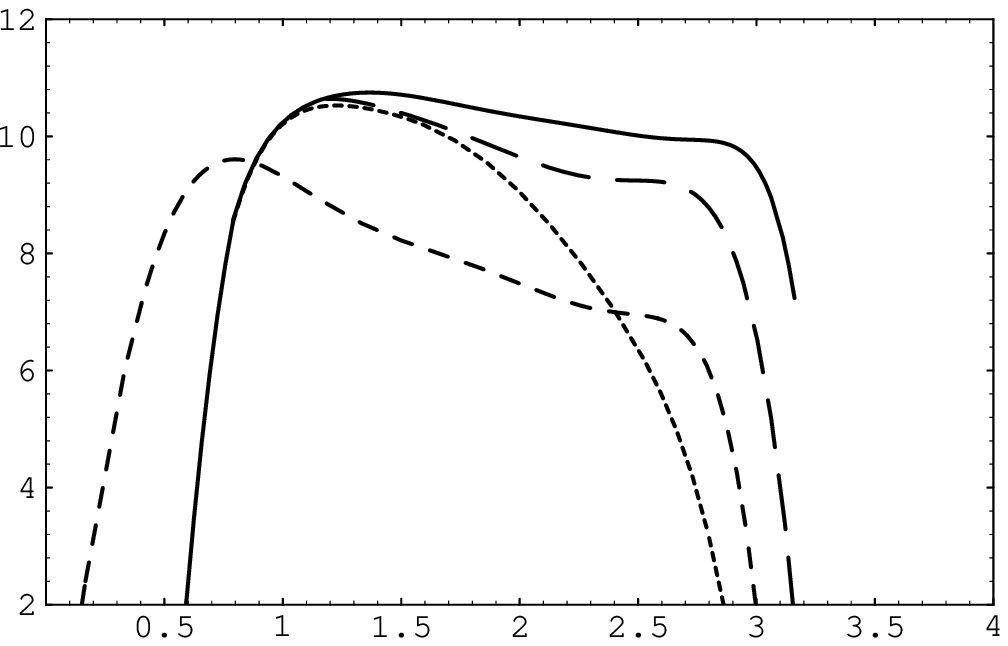}
    }
  \vspace*{-1.5cm}
  \noindent
  \caption{Results of the LME for the real finite parts
    of diagrams 7.1 and 7.2 ~.
    Solid curves represent the small-$q^2$ expansions,
    middle-dashed the leading term of the LME,
    short-dashed the sum of 11 terms in the LME,
    long-dashed the [5/5] Pad\'e approximant from the LME.}
\end{figure}
                                %\end{center}

                                %Figure 4: Results of the LME for the real finite parts 
                                %of diagrams 7.1 and 7.2 ~.
                                %Solid curves represent the small-$q^2$ expansions, 
                                %middle-dashed the leading term of the LME,
                                %short-dashed the sum of 11 terms in the LME,
                                %long-dashed the [5/5] Pad\'e approximant from the LME.

Finally the LME of the above diagrams has the
following general form:

\begin{equation}
  F_{\rm as}^N = \frac{1}{m_t^4}
  \sum_{n=-1}^N \sum_{i,j=-1 \atop i+j=n}^n 
  \left( \frac{M_W^2}{m_t^2} \right)^i \left( \frac{q^2}{m_t^2} \right)^j
    \sum_{k=0}^m A_{i,j,k}(q^2,M_W^2,{\mu}^2) \ln^k \frac{m_t^2}{\mu^2}
    \label{series}
  \end{equation}
  where $m$ is the highest degree of divergence (ultraviolet, infrared, collinear)
  in the various contributions to the LME ($m\le$ 3 in the 
  cases considered). 
  $M_W^2/m_t^2$ and $q^2/m_t^2$ are considered as small parameters. 
  $A_{i,j,k}$ are in general complicated functions of the arguments, i.e. they
  may contain logarithms and higher polylogarithms.

  In the next Section we present results and compare the two methods described
  here.

  \section{Results}

  The small momentum expansion of cases 7, 7.1 and 7.2 is described
  in detail in Ref. \cite{Mainz}. The additional 
  subgraph arising in these cases 
  is the same irrespectively of the mass distribution
  and is shown in Fig.5 ($case$ 7).   

  In the case of the LME two additional subgraphs arise in
  each of the cases 7.1 and 7.2. Beyond that these sets of additional
  subgraphs are also of quite different structure. Furthermore the $2^{\rm nd}$ 
  and $4^{\rm th}$ 
  graphs in the row (see Fig.5) produce $1/\epsilon^3$
  terms which cancel, however. Since there are no UV divergences, these must
  be mixtures of infrared and collinear ones. 
  They determine the highest degree of divergence in these
  cases and thus the highest power of the logarithm as discussed in (4). 

  Our numerical results for cases 7.1 and 7.2 are presented in Fig.6,
  and for $q^2=M_Z^2$ in Table 1. In the figures we show the small 
  $q^2$-expansion (solid line) in comparison with the lowest order approximation
  (middle-dashed) and the sum of terms (small-dashed) with $N$=9, see (4). 
  The scale
  parameter $\mu=m_t$, i.e. only $k$=0 contributes in (4). We see that up to 
  $q^2=M_Z^2$ the sum of 11 terms agrees quite well with the 
  small $q^2$-expansion,
  while for higher $q^2$ the agreement very quickly worsens. For this reason we
  formally apply also the Pad\'{e} summation technique. With 11
  terms in the series, a [5/5]-approximant (long-dashed) can be constructed. It
  is seen that indeed this improves the situation considerably 
  up to $q^2=4 M_Z^2$
  though for a better agreement many more terms in the LME would be
  needed. In the small $q^2$-expansion only 9 terms were taken into account, i.e.
  a [4/4] approximant is calculated. From Table 1 we see that for $q^2=M_Z^2$ 
  indeed a rather precise 
  result can be obtained with 11 terms from the LME, in particular
  if Pad\'e's are applied \cite{FLM}.

  $F^{(0)}$ in Table 1 corresponds to the lowest order of the diagrams
  (N=-1 or 0, respectively) and $F^{(4)}$ to the order of expansion performed in
  Ref. \cite{HSS}.

                                %
                                %   TABLES
                                %

  {\scriptsize
    \begin{table}[h]
      \caption{Values of diagrams at $q^2=90^2,\,
        \mu=m_t=180,\, m_W=80$.}

      \begin{tabular}{|l|ll|ll|} \hline
        & case 7.1  && case 7.2 &  \\ \hline
        $F^{(0)}$  
        & $6.4$       & $-i30.0$
        & $9.2$       & $-i24.5$ \\
        $F^{(4)}$ 
        & $11.1$       & $-i17.93$ 
        & $9.4$     & $-i44.848$  \\
        $F^{(10)}$ 
        & $10.1$       & $-i17.95218$ 
        & $10.122$     & $-i44.84523$  \\
        $[5/5]$    
        & $9.996$       & $-i17.9528$   
        & $10.189$      & $-i44.842$  \\
        small-$q$  
        & $9.992668259$  & $-i17.95215366$  
        & $10.193902102$  & $-i44.845273975$  \\ \hline
      \end{tabular}
      \label{LME1}
    \end{table}
    }

  \section{Bremsstrahlung}

  Of particular interest is the calculation of Bremsstrahlung
  in terms of the LME. Let us consider again the decay $Z \to b\bar{b}$.
  The kinematics of the (gluon-) Bremsstrahlung for this process
  is given by

  $$
  q\to p_1-p_2+p_3,\qquad  p_1^2=p_2^2=p_3^2=0.
  $$
  Thus we have 3 invariants
  $$
  p_1p_2,\qquad p_2p_3,\qquad p_3p_1.
  $$

  We are interested in the integrated bremsstrahlung.
  The phase space measure in $d$ dimensions can be
  written as
  \begin{eqnarray}
    \int d\Gamma_3 &=& \int \delta^{(d)}(q - p_1+p_2-p_3)
    \Pi_{i=1}^{3}
    \frac{d^{d-1} \bf p_i}{(2\pi)^{d-1}2p_{i0}} \nonumber\\
    &=&
    \frac{1}{(2\pi)^{3d-3}} \frac{2(2\pi)^{d-2}}{\Gamma(d-2)}
    \frac{q^2}{32} \left(\frac{q^2}{2}\right)^{d-4} \nonumber\\
    && \int_0^1 dx\,dy\,
    x^{d-3} (1-x)^{d/2-2} y^{d/2-2} (1-y)^{d/2-2}
  \end{eqnarray}
  and the invariants can be expressed in terms of $x,y$ as
  \begin{eqnarray}
    p_1p_2 &=& \frac{q^2}{2}x(1-y)\,,  \nonumber\\
    p_2p_3 &=& \frac{q^2}{2}(1-x)\,,   \nonumber\\
    p_3p_1 &=& \frac{q^2}{2}xy   \,.   \nonumber
  \end{eqnarray}

  For the 1-loop amplitude we apply the LME
  w.r.t the top mass $m_t^2$. For the diagrams with $M_W$ and $m_t$
  mass in virtual lines the desired expansion looks like
  \begin{eqnarray}
    A = \sum_{N}\sum_{i+j+k+n=N}
    \left(\frac{p_1p_2}{m_t^2}\right)^i
      \left(\frac{p_2p_3}{m_t^2}\right)^j
        \left(\frac{p_3p_1}{m_t^2}\right)^k
          \left(\frac{m_W^2}{m_t^2}\right)^n
            f_{ijkn}\,,
          \end{eqnarray}
          where $f_{ijkn}$ are cubic polynoms of $\log(m_t^2/\mu^2)$
          with coefficients being  functions of
          $q^2,m_W^2,\mu^2$, i.e.
          $$
          f_{ijkn} =
          a \log^3\frac{m_t^2}{\mu^2}
          +b \log^2\frac{m_t^2}{\mu^2}
          +c \log\frac{m_t^2}{\mu^2}
          +d.
          $$

          After expansion the integration over the phase space
          is trivial (it can be done completely in terms of $\Gamma$-functions).

          \section{Diagrams with only one non-zero mass }

          So far the method of expansion is considered
          as semianalytic in the sence that only a limited
          number of coefficients can be obtained explicitely.
          In this Section, however, we want to go one step further.
          We calculate the first few coefficients of the
          expansion of a diagram with only one non-zero mass
          by means of the LME \cite{LME}       
          method. The method of differential equations \cite{DEM}
          then yields an idea, like in \cite{FKV}, what the general
          analytic form of these coefficients might be, providing
          some ``basis'' in terms of which they might be expressed.
          The Ansatz of equating the explicit coefficients obtained
          from the LME to a linear combination
          in terms of the basis elements, yields a system of linear
          equations, which can be solved to yield the desired representation
          of the coefficients.

          The main problem in this approach is the choice of the
          basis elements. We start from so-called harmonic sums
          which are particularly relevant for moments of
          structure functions in QCD (e.g. \cite{Yndurain}). 
          These functions are directly related to
          (generalized) polylogarithms \cite{polylogarithms}
          and therefore it is not surprising that they appear
          in the analysis of massive diagrams.
          However, not every massive
          diagram even of self-energy type can be expressed
          in terms of these sums. Apart from harmonic sums we introduce a new
          type of sums which we call $W$- and $V$-sums.

          Our starting point is the LME \cite{LME}
          of the diagrams. For the particular diagrams under consideration
          it was discussed in detail in \cite{ft}.
          Here we just note that the result of a large mass expansion
          for these diagrams reads ($z=q^2/m^2$)
          \begin{equation}\label{lme}
            J = \frac{1}{(q^2)^a}\sum_{n\geq 1} z^n
            \sum_{j=0}^\omega \frac{1}{\varepsilon^j}
            \sum_{k=0}^\nu \log^k(-z)
            A_{n,j,k}\,,
          \end{equation}
          where $a$ is the dimensionality of the diagram,
          $\omega$ and $\nu$ independent of $n$ are
          the highest degree of divergence and
          the highest power of $\log(-q^2/m^2)$, respectively
          (in our cases $\omega,\nu\leq4$).
          The coefficients $A_{n,j,k}$ are of the form
          $r_1+\zeta_2 r_2+\dots+\zeta_\nu r_\nu$ with
          $r_c$ being rational numbers and $\zeta_c=\zeta(c)$ is the
          Riemann $\zeta$-function.

          Series (\ref{lme}) always has a nonzero radius of
          convergence, which is defined by the position
          of the nearest nonzero threshold in the $q^2$-channel.
          For brevity we shall call $m$-cuts ($2m$-cuts, etc.)
          possible cuts of a diagram in $q^2$ corresponding
          to 1 (2, etc.) massive particles in the intermediate state.

          We start from diagrams with the simplest threshold
          structure, i.e. with $m$-cuts and $0$-cuts.
          It turns out that all of these with only one massive line
          are expressible in terms of
          harmonic sums $S_k(n-1)=\sum_{j=1}^{n-1} 1/j^k$
          and alternating harmonic sums $K_a(n-1)=\sum_{j=1}^{n-1}(-)^{j+1}/j^a$.

          As a first example consider the 2-loop 2-point function
          of Fig.~7, $J_{(a)}$, which was considered in \cite{BroadhurstZP47}.
          Using the standard large mass
          expansion technique one can get the first few coefficients of the
          expansion of this diagram in powers of $z=q^2/m^2$
          \begin{equation}\label{sample1}
            \frac{1}{q^2}\sum_{n=1}^{\infty} a_n z^n
            = \frac{1}{q^2}\Biggl(
            2\zeta_2 z
            + \left( \zeta_2 + \frac12 \right) z^2
            + \left( \frac23\zeta_2 + \frac12 \right) z^3
            + \dots \Biggr)\,,
          \end{equation}
          where $\zeta_2=\zeta(2)$ is the Riemann $\zeta$-function.

          \begin{figure}[h]
            \centerline{\vbox{\epsfysize=30mm \epsfbox{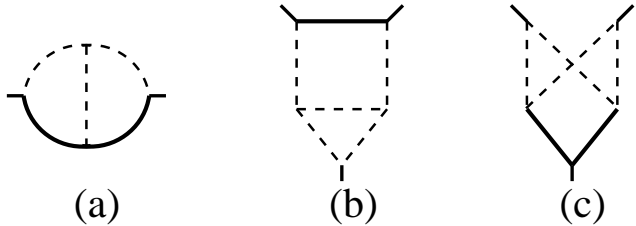}}}
\vspace*{2mm}
            \caption{\label{Fig_3} Three diagrams, selfenergy and vertices, 
              for the TE of which results are reported.}
          \end{figure}

          Our first step is to find an expression for
          the higher order terms of the series (\ref{sample1}) as
          a functions of $n$ i.e. $a_n=a(n)$.
          
          To achieve this let us search for $a_n$ as a linear
          combination of harmonic sums.
          The rule is that one has to
          take into account all possible products of the type
          $\zeta_a S_{b_1}\dots S_{b_k}/n^c$ with
          the 'transcendentality level' (TL) $a+b_1+\dots+b_k+c=3$. 
          It is obvious that one
          can exclude $\zeta_3$ from the very beginning since it never 
          appears on the r.h.s. of (\ref{sample1}).
          Thus we have the following Ansatz for $a_n$
          \begin{equation}\label{ansatz}
            a_n = \frac{\zeta_2}{n}x_1 + \zeta_2 S_1 x_2
            + S_3 x_3 + S_2 S_1 x_4 + S_1^3 x_5
            + \frac{S_2}{n} x_6 + \frac{S_1^2}{n} x_7
            + \frac{S_1}{n^2} x^8 + \frac{1}{n^3} x_9\,,
          \end{equation}
          where $x_1,\dots,x_9$ are rational numbers independent
          on $n$. The argument of the $S$-functions is $n-1$.
We refer to the structures in (\ref{ansatz})
as `basis elements'. Indeed, the functions
$\zeta_aS_{b_1}(n-1)\dots S_{b_k}(n-1)/n^c$ are
algebraically independent.

  Inserting the expression for $a_n$ from (\ref{ansatz})
into the l.h.s. of (\ref{sample1})
and equating equal powers of $z$, we obtain
a system of linear equations for the $x_i$.
An explicit computation ensures that the system can be solved
in terms of rational numbers for the $x_i$
The solution is $x_1=2,x_6=2,x_8=-2,x_{2,3,4,5,7,8}=0$ i.e.
the answer for the diagram at hand is
\begin{equation}\label{res1}
J_{(a)} =
  \frac{1}{q^2}
  \sum_{n=1}^{\infty} z^{n}
       \Biggl(
     2 \frac{\zeta_2}{n} + 2 \frac{S_2(n-1)}{n} - 2\frac{S_1(n-1)}{n^2}
       \Biggr)\,.
\end{equation}

  Equation (\ref{ansatz}) can be considered as expansion of
the general coefficient $a_n$ in terms of a `basis'' with
(unknown, rational) coefficients $x_i$. Now the question arises:
what are the basis elements in general?
So far we are lacking rules for predicting a basis of a given diagram.
The power of the method, however, is that given a set of basis
elements for one diagram, it can be used to find the solution
for other diagrams. Often, though, one has to `generalize'
already known (harmonic) sums in a basis.

As an illustration
consider the diagram shown in Fig.~7, $J_{(b)}$. Its large mass expansion
looks like
\begin{equation}\label{sample2}
  J_{(b)} =
   \frac{1}{(q^2)^2} \sum_{n=1}^\infty
     z^n \Biggl(
       r_n^{(2)} \log^2(-z)
     + r_n^{(1)} \log(-z)
     + r_n^{(0,3)} \zeta_3 + r_n^{(0,2)} \zeta_2
     + r_n^{(0,0)}
                         \Biggr)\,,
\end{equation}
with the $r$'s being rational numbers. It is obvious that
one can search for a solution for each of the $r$'s independently.
Again we use the same set of functions ($1/n^a$ and $S_b$) as above
but now with different transcendentality level(s).
The system of equations has a solution only if we add
the factor $(-)^n$ which can be seen easily
by inspection of the series. At the end we arrive at
\begin{eqnarray}\label{res2}
  J_{(b)} &=&
   \frac{1}{(q^2)^2} \sum_{n=1}^\infty
     z^n (-)^n \Biggl(
       \frac{S_1}{n} \log^2(-z)
     + \Bigl( - 4 \frac{S_2}{n} + \frac{S_1^2}{n}
      - 2 \frac{S_1}{n^2} \Bigr) \log(-z)
                    \nonumber\\
  &&  - 6 \frac{\zeta_3}{n} + 2 \frac{\zeta_2 S_1}{n}
     + 6 \frac{S_3}{n} - 2 \frac{S_2 S_1}{n}
     + 4\frac{S_2}{n^2} - \frac{S_1^2}{n^2}
     + 2\frac{S_1}{n^3}
                         \Biggr)\,,
\end{eqnarray}
where as above we take all harmonic sums ($S_i$) with the
argument $n-1$, which is omitted.

We point out that in (\ref{res2})
the part without $\log(-z)$ has basis elements
$\zeta_aS_b/n^c$ obeying $a+b+c=4$ (i.e. it has a basis
with TL$=4$). Terms proportional to $\log(-z)$ are of 3rd level
while those proportional to $\log^2(-z)$ of the 2nd.
One can say that $\log(-z)$ itself is of 1st level
and each $\log^a(-z)$
reduces the level of basis elements by $a$ units.
This is the general behaviour for all diagrams we consider.
The same rule applies to UV and IR poles $1/\varepsilon$
if they are present in a diagram i.e. each $1/\varepsilon^a$
reduces the level of basis elements by $a$ units.

We also stress the presence of the
factor $(-)^n$. Strictly speaking the basis now is not
$\zeta_aS_b/n^c$ but rather $(-)^n\zeta_aS_b/n^c$.

Finally we observe the following recipe of generalization,i.e.
elements of higher TL ($a+b$) can be constructed in the form:

\begin{eqnarray}
  S_{a,b}(n) &=& \sum_{j=1}^{n} \frac{1}{j^a} S_b(j-1) \,,      \\
  K_{a,b}(n) &=& - \sum_{j=1}^{n} \frac{(-)^j}{j^a} S_b(j-1)\,. \\
\end{eqnarray}

  The structure of the lowest level
for diagrams having both $m$- and $2m$-cuts
is more complicated. We find it from the nonplanar graph shown in
Fig.~7, $J_{(c)}$. Namely this diagram has
a double collinear pole and we
can easily perform the integrations in calculating the
$1/\varepsilon^2$ contribution with the result

\begin{eqnarray}
\label{polepart}
J_{(c)} & \stackrel{1/\varepsilon^2}{=} &
  - \frac{1}{\varepsilon^2 (q^2)^2}\int_0^1
  \frac{d\alpha_1\, d\alpha_2}
  {(1-\alpha_1(1-\alpha_2)z) (1-(1-\alpha_1)\alpha_2 z)} \nonumber\\
  &=&
   -\frac{1}{\varepsilon^2 (q^2)^2}
    \sum_{n=1}^\infty z^n \frac{(n!)^2}{(2n-1)!}
    3\sum_{j=1}^{n-1} \frac{(2j)!}{(j!)^2} \frac{1}{j} \nonumber\\
  &=&
   -\frac{1}{\varepsilon^2 (q^2)^2}
    \sum_{n=1}^\infty z^n {2n\choose n}^{\!\!\!-1} 6\,\frac{W_1(n-1)}{n}.
\end{eqnarray}

If we assign to
the factor ${2n\choose n}^{\!\!\!-1}$ the 0th level,
then the expression on the r.h.s of (\ref{polepart}) has
the correct 2nd level as it should be for the double
pole part of a vertex function. The generalization is done according
to

\begin{eqnarray}
  W_{a,b}(n) &=& \sum_{j=1}^{n}
           {2j\choose j} \frac{1}{j^a} S_b(j-1)\,.                \\
\end{eqnarray}

The result for the diagram under consideration finally is

\begin{eqnarray}
%
%  case N12
%
J_{(c)} &=&
   \frac{1}{(q^2)^2}
    \sum_{n=1}^{\infty} z^n {2n\choose n}^{\!\!\!-1}\frac{1}{n} \nonumber\\
  &&  \Biggl\{
    - \frac{6}{\varepsilon^2} W_1
    +\frac{1}{\varepsilon} \Bigl[
          - 4 W_2
          - 12 W_{1,1}
          - 12 S_1 W_1
          - 16 S_2
               \Bigr]  \nonumber\\
 &&
      - 18 \zeta_2 W_1
      - 4 W_3
      - 8 S_3
      + 12 W_{1,2}
      - 8 W_{2,1}
      - 12 W_{1,(1+1)}  \nonumber\\
 &&
      - 16 S_{1,2}
      - 8 S_1 W_2
      - 24 S_1 W_{1,1}
      - 16 S_1 S_2
      - 12 S_1^2 W_1
      \Biggr\},
\end{eqnarray}

where

\begin{eqnarray}
  S_{a,(b+c)}(n) &=& \sum_{j=1}^{n}
      \frac{1}{j^a} S_b(j-1) S_c(j-1)\,,                    \\
  W_{a,(b+c)}(n) &=& \sum_{j=1}^{n}
      {2j\choose j} \frac{1}{j^a} S_b(j-1) S_c(j-1)\, ~etc.,
\end{eqnarray}
all have TL$=a+b+c$.

   For completeness we mention a further type of basis elements
which we discovered from the application of the differential
equation method \cite{FKV}

\begin{equation}
  V_a(n-1) = \sum_{j=1}^{n-1} {2j\choose j}^{\!\!\!-1} \frac{1}{j^a}.
\end{equation}

   Further basis elements were obtained in a similar fashion. The
result for a great variety of diagrams will be published elsewhere.

\section{Conclusion}

   The purpose of many of our investigations is to test various methods for
the calculation of two-loop diagrams needed for the evaluation of
measured
processes in the full SM of the electroweak theory. Comparing
different methods yields information about their precision and
possible applicability. In particular we have forced the developement
of packages for the large mass expansion and found that in the
domain of interest for the processes under consideration this
method can be used with reasonable liability. For the case of only
one non-zero mass a completely new approach was found, namely 
``guess and verify'', to set up analytic expressions for many diagrams
of interest.

\newpage


\begin{thebibliography}{99}

\bibitem{ZT}
J. Fleischer, O.V. Tarasov, F. Jegerlehner and
 P.~R\c{a}czka, Phys. Lett. B293 (1992) 437;
G. Buchalla and A.J. Buras, Nucl. Phys. B398
 (1993) 285;
G. Degrassi, Nucl. Phys. B407 (1993) 271;
K.G. Chetyrkin, A. Kwiatkowski and M. Steinhauser,
 Mod. Phys. Lett. A8 (1993) 2785.

\bibitem{HSS} R.Harlander, T.Seidensticker and M.Steinhauser, hep-ph/9712228,
preprint MPI/PhT/97-81, TTP97-52;

\bibitem{FLM} J. Fleischer, M. Kalmykov and O. Veretin,
Phys. Lett. B427 (1998) 141.

\bibitem{TLAMM}
L.V.~Avdeev, J.~Fleischer, M.~Yu.~Kalmykov, M.~Tentyukov,
Nucl.Instrum. Meth. A 389 (1997) 343;
{\it Towards Automatic analytic Evaluation of Diagrams with Masses},
Comp.Phys.Comm. 107 (1997) 155.

\bibitem{LME}
F.V.~Tkachov, Preprint INR P-0332, Moscow (1983); P-0358, Moscow (1984);
~~K.G.~Chetyrkin,
Teor. Math. Phys. 75 (1988), 26; ibid 76 (1988), 207;
Preprint, MPI-PAE/PTh-13/91, Munich (1991);
~~V.A.~Smirnov,
Comm. Math. Phys. 134 (1990), 109;
{\it Renormalization and asymptotic expansions}
(Birkh\"auser, Basel, 1991).

\bibitem{DIA}
L.V.~Avdeev, J.~Fleischer, M.~Yu.~Kalmykov, M.~Tentyukov,
Nucl.Instrum. Meth. A 389 (1997) 343;
{\it Towards Automatic analytic Evaluation of Diagrams with Masses},
accepted for publication in Comp.Phys.Comm.,(hep-ph/9710222);
L.V.~Avdeev, M.Yu.~Kalmykov, Nucl. Phys. B502 (1997) 419;
J. Fleischer and M. Tentukov, {\it A Feynman Diagram Analyser DIANA},
in preparation.

\bibitem{QGRAF}
P.~Nogueira, J. Comput. Phys. 105 (1993), 279.

\bibitem{FORM}
J.A.M.~Vermaseren: Symbolic manipulation with
FORM, Amsterdam, Computer Algebra Nederland, 1991.

\bibitem{ft}
J.~Fleischer and O.V.~Tarasov,
{  Z.Phys.}, {\bf C 64} (1994) 413;
J.Fleischer and O.V.Tarasov, in proceedings of the ZiF conference on
{ \elevenit Computer Algebra in Science and Engineering},
Bielefeld, 28-31 August 1994, World Scientific 1995, J. Fleischer,
J. Grabmeier, F.W.Hehl and W.K\"uchlin editors.
J. Fleischer, V. A. Smirnov and O. V. Tarasov,
Z.Phys.{\bf C74} (1997) 379.

\bibitem{Recur} A.I.~Davydychev and J.B.~Tausk,
{ Nucl.~Phys.},        {\bf B397} (1993) 123.

\bibitem{bft} D.J.~Broadhurst, J.~Fleischer and O.V.~Tarasov,
{  Z.Phys.}, {\bf C 60} (1993) 287.

\bibitem{davt} A.I.~Davydychev and J.B.~Tausk,
{ Nucl.~Phys.}, {\bf B465} (1996) 507.

\bibitem{Tara} O.V.~Tarasov,
{ Nucl.~Phys.},        {\bf B480} (1996) 397.

\bibitem{JF}
J. Fleischer,
Int.J.Mod.Phys.{\bf C6} (1995) 495;

\bibitem{Mainz}
J.~Fleischer et al., Eur.Phys.J. C2 (1998) 747.

\bibitem{Sha} D.~Shanks, { J.~Math.~Phys.} {\bf 34} (1955);
P.~Wynn, { Math.~Comp.} {\bf 15} (1961) 151; G.A.~Baker,
P.~Graves-Morris, Pad\'e approximants, in { Encyl.of math.and its
appl.,} Vol. {\bf 13, 14}, pp Addison-Wesley (1981).

\bibitem{BGW} G.A.~Baker, Jr., J.L.~Gammel and J.G.~Wills,
{ J.Math.Anal.Appl.}, {\bf 2} (1961) 405;
G.A.~ Baker, Jr.,
{ Essentials of Pad\'e Approximants}, pp Academic Press (1975).

\bibitem{DEM}
A.V. Kotikov, Phys.Lett. {\bf B254} (1991) 185;
ibid. {\bf B259} (1991) 314; ibid. {\bf B267} (1991) 123.

\bibitem{FKV} J.~Fleischer, A.V.~Kotikov and O.L.~Veretin,
Phys.Lett. {\bf B417} (1998) 163.

\bibitem{Yndurain}
A. Gonz\'alez-Arroyo, C. L\'opez and F.J. Yndur\'ain,
Nucl. Phys. B153 (1979) 161;
J.A.M. Vermaseren, NIKHEF-98-14, hep-ph/9806280.

\bibitem{polylogarithms}
A.Devoto and D.W.Duke, La Rivista del Nuovo Cimento, Vol. 7, No. 6 (1984) 1;
K.S. K\"olbig, SIAM Journal on Mathematical Analysis,
Vol. 17, No. 5 (1986) 1232.

\bibitem{BroadhurstZP47}
D.J. ~Broadhurst, Z.Phys. {\bf C47  } (1990) 115.

\end{thebibliography}
\end{document}